# TOWARDS COMBATING PANDEMIC-RELATED MISINFORMATION IN SOCIAL MEDIA

Isa Inuwa-Dutse

*University of St Andrews*

*[Iid1@st-andrews.ac.uk](mailto:Iid1@st-andrews.ac.uk)*

## ABSTRACT

*Conventional preventive measures during pandemic include social distancing and lockdown. Such measures in the time of social media brought about a new set of challenges - vulnerability to the toxic impact of online misinformation is high. A case in point is the prevailing COVID-19; as the virus propagates, so does the associated misinformation and fake news about it leading to infodemic. Since the outbreak, there has been a surge of studies investigating various aspects of the pandemic. Of interest to this chapter[1] include studies centring on datasets from online social media platforms where the bulk of the public discourse happen. Consequently, the main goal is to support the fight against negative infodemic by (1) contributing a diverse set of curated relevant datasets (2) recommending relevant areas to study using the datasets (3) discussion on how relevant datasets, strategies and state-of-the-art IT tools can be leveraged in managing the pandemic.*

Keywords: Coronavirus, SARS-CoV-2, COVID-19 Datasets, Infodemic, Information Diffusion, Online Social Networks, Twitter,

## INTRODUCTION

Human history is intertwined with various pandemics, infectious disease on a global scale, events resulting in a dramatic high mortality rate and economic hardship. Pandemics from diseases such as smallpox, tuberculosis, and the Spanish flu resulted in a large number of lost lives (Kaur, 2020). Recently, one of the defining moments of the year 2020 is the outbreak of the zoonotic Coronavirus Disease (COVID-19) that radically disrupts normal social interactions. The virus was first reported by the World Health Organisation (WHO) on December 31, 2019, in Wuhan, China. At the time of writing this chapter, the recent statistics from the WHO reported 108,153,741 confirmed cases and 2,381,295 confirmed deaths across 223 countries, areas or territories. It is easy to be oblivious of early warnings despite apparent reasons suggesting otherwise. When the prevailing pandemic was first reported, many nations were heedless in taking proactive measures to a point that the outbreak quickly overwhelmed healthcare facilities making it difficult to attend to ailing people, fatigue from health workers, distress and grieve from families of ailing and lost ones. The lacklustre attitude from some leaders and the politicisation of the pandemic further compounds the situation, resulting in confusing and conflicting narratives. As the pandemic crisis exacerbates, many forms of preventive and curative responses have been imposed to curtail the scale of spread and negative impact of the pandemic. Figure 1 shows a summary of the total number of cases globally[2].

---

[1] To appear in the book 'Data Science Advancements in Pandemic and Outbreak Management', IGI Global Acquisitions

[2] The data used in plotting the figure is obtained from [www.worldometers.info](http://www.worldometers.info)

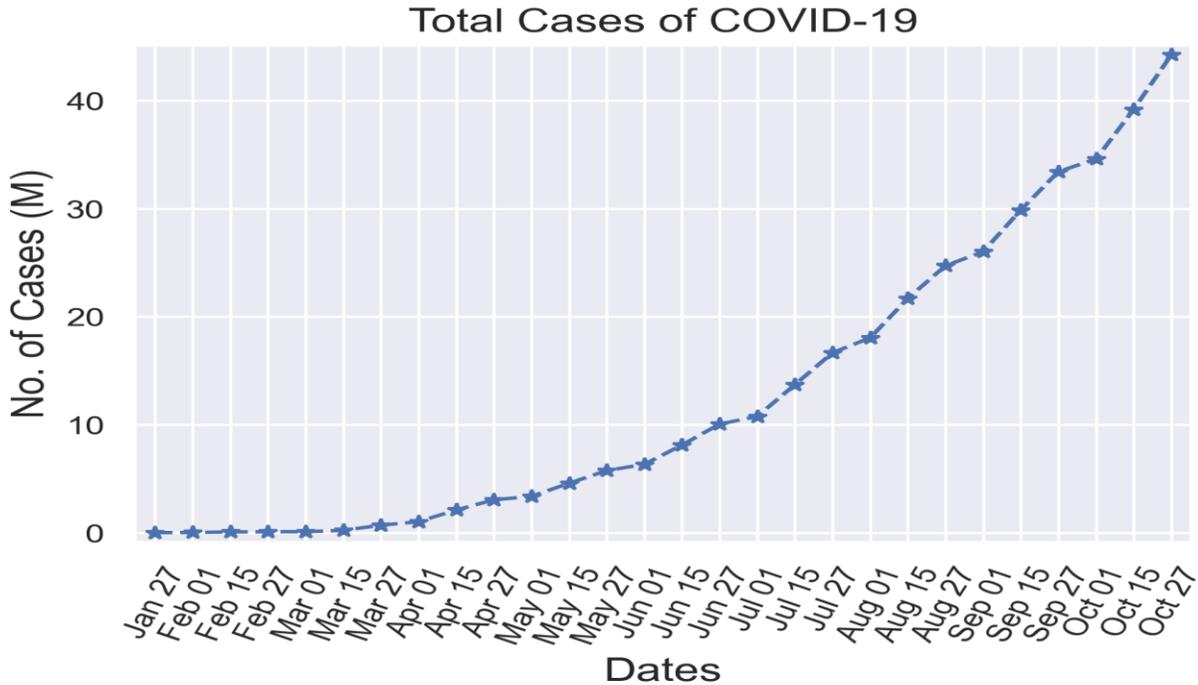

*Figure 1. A summary of the total number of COVID-19 global cases.*

Various reactionary approaches, sometimes impulsive, have been used to flatten infection peaks to avoid overwhelming the prevailing healthcare facilities and alleviating the associated financial challenges. Typical measures to slow down the infection rate include disinfection, social distancing, contact tracing, isolation/quarantine and some curative measures. Following the traditional approach of mitigating spread, the infamous *lockdown* measure introduced to curb the virus spread has altered many aspects of social routines in which demand for online-based services skyrocketed. While modern-day online social media networks, such as Facebook[3] and Twitter[4], facilitate the spread of information to a wide audience, making it a useful facility for instant information updates and socialisation, they also present new sets of challenges. Figure 2 shows a summary of the major events[5] since the outbreak.

With a substantial proportion of the populace confined to their homes for a long period, vulnerability to the toxic impact of online misinformation is high during the COVID-19 outbreak. There is a growing body of work tackling many problems associated with the outbreak. For instance, concerning infodemic[6], coined to denote the online outburst of pandemic-related information, especially misinformation, researchers have been curating and documenting various datasets about COVID-19 pandemic. Of interest to this chapter are studies centring on online datasets, with emphasis on the online social media platforms where the bulk of the public discourse happen, regarding spurious content associated with the pandemic. Ultimately, the chapter's goal is to support the fight against online misinformation with particular emphasis on pandemic-related datasets.

---

[3] https://www.facebook.com/
[4] https://twitter.com/
[5] For a more comprehensive information about the major events see
https://www.whoint/emergencies/diseases/novel-coronavirus-2019/interactive-timeline
[6] See https://www.who.int/docs/default-source/coronaviruse/situation-reports/20200202-sitrep-13-ncov-v3.pdf?sfvrsn=195f4010_6

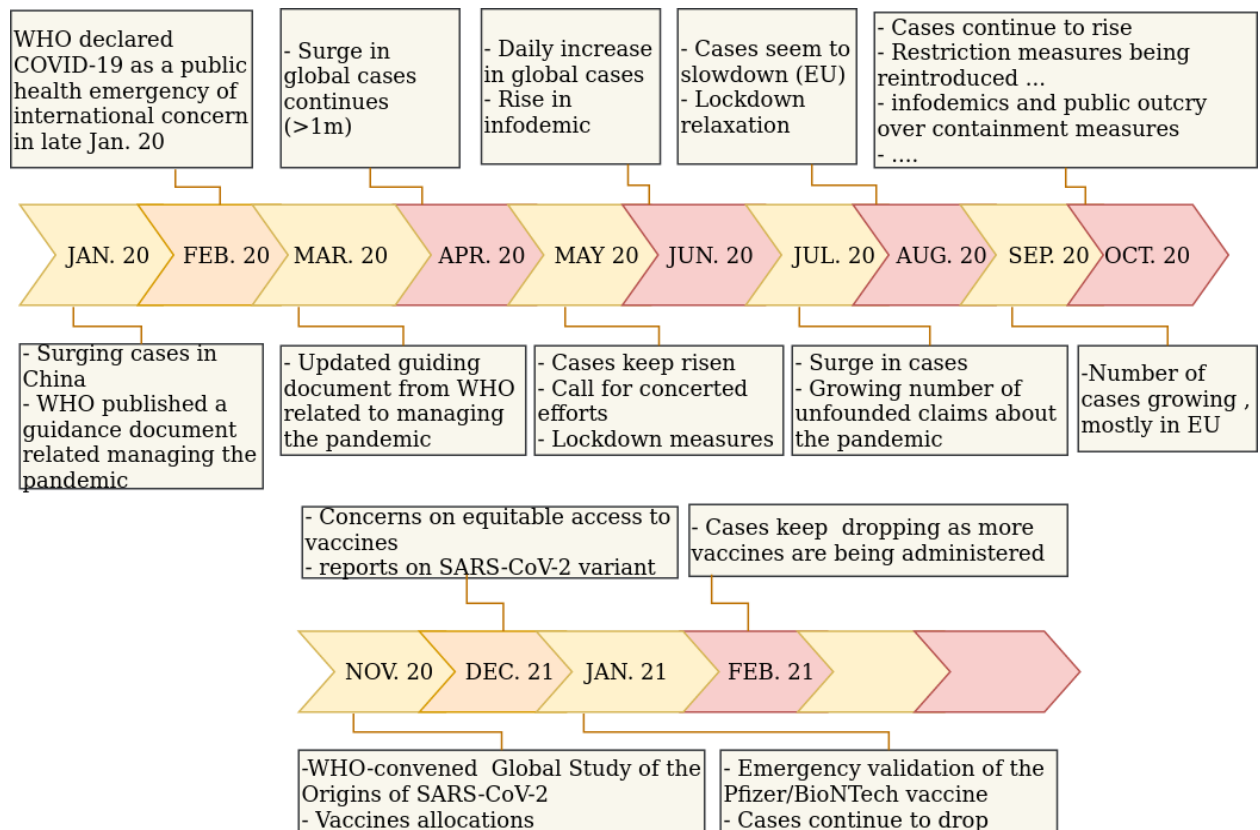

*Figure 2. A chronicle of some major events or topical issues related to the management of the pandemic.*

Many aspects of the pandemic can be explored using such data collection – from leveraging benchmarking datasets to assess the veracity of information related to the pandemic (infodemic) to the more advanced task of modelling and tracing the propagation of the virus. Consequently, the chapter will, among other benefits, better inform how to ensure that relevant pandemic-related content dominates, and irrelevant content is suppressed, especially during critical times of the pandemic. Noting the growing vulnerability to the toxic impact of online misinformation during the pandemic, the contributions of the chapter include:

– ***Datasets to neutralise negative infodemic:*** the study presents a diverse set of curated relevant datasets about the pandemic to help in curtailing the proliferation of harmful information about the outbreak. The datasets will support studies investigating various aspects of the pandemic through the prism of online social media platforms.
– ***How to leverage the datasets for actionable insights:*** the ultimate goal is to support the fight against online misinformation on pandemic-related issues, hence, supplementing the contributed datasets with a discourse on some relevant actionable areas to focus on will be beneficial. Relevant tasks to undertake include leveraging benchmarking datasets to assess the veracity of information related to the pandemic, modelling and tracing the propagation of the virus from online signals.

The chapter is structured as follows: Part I offers relevant background information regarding pandemics and Part II presents a detailed account of the relevant datasets, including the collection and processing steps, to use. Part III proffers some research problems worthy of investigation using the datasets. Finally, Part IV concludes the study with a closing remark.

## BACKGROUND

As pointed out earlier, some of the swift measures taken to mitigate pandemics include disinfection, social distancing, contact tracing and *lockdown*. Moreover, palliatives are being provided to cushion the financial hardships brought about by the pandemic. With technological advancements, all the above-mentioned measures can be improved. Therefore, the focus in this section is on existing related work, and to highlight relevant topics that would help in achieving such goals, notably from an online social media perspective.

**Pandemic in the Age of Online Connectivity**

Arguably, the need for online-based services has never been in higher demand than what is being witnessed in the COVID-19-induced lockdown era. One of the implications of the lockdown is the relegation of virtually all human engagements to the online realm, which also results in an increasing number of uncensored posts. Noting how misleading information can have catastrophic consequences and hampers the fight about applying containment measures, it is pertinent to combat the pandemic from all possible fronts. While the architecture of online social media networks simplifies the spread of information to a wide audience, it also enables a breeding ground for misleading information. This opens another frontier of challenges in the fight against the virus.

Drawing from *ethnography*, a form of social research that is concerned with individual culture and group behaviour, *netnography* is a relatively new term coined to denote the use of online social media platforms to study people's interactions and behaviours (Kozinets, 2007; Pink, 2016). It encompasses aspects of data collection, analysis, research ethics, and representation, rooted in participants' observations. Because observational data can be retrieved from online communities or groups, effective aggregation of such data would yield useful insight (Kozinets, 2007). It can be argued that modern-day social interactions will be incomplete without taking online social relationships into account, where various forms of interactions among diverse users happen. This capability makes it possible to empirically quantify and evaluate social relationships among users on an unprecedented scale. Essentially, many social network theories and analytical solutions can now be tested using real social media data. Thus, a treatise on contemporary social engagements will be incomplete without reference to online social networks.

*Online Social Media Ecosystem*

The transformative power of technological advancements across various facets of public lives is quite enormous. Many forms of social interactions are continually evolving to support a myriad of objects to remain connected through a communication model that enables a multi-flow of information (see the influence network model of Watts and Dodds (2007)), thus contrasting it with the two-step flow model in which few users mediate communication between the media and the general public (Katz et al., 2017).

The contemporary social media ecosystem consists of numerous platforms which support various aspects of humans' social engagements and enable users to simultaneously generate and consume information (Inuwa-Dutse et al., 2020). Many forms of social interactions are continually evolving through platforms such as Twitter, Facebook, WhatsApp, TikTok, LinkedIn, Snapchat, Twitch, Pinterest, YouTube, Viber, that facilitate information diffusion and socialisation at scale. These platforms have been instrumental in socialisation, breaking news, globalisation and enabling socio-technological research (Sundaram et al. 2012). Moreover, the platforms are quite popular with the public; thus, it is worthwhile understanding the social media ecosystem in great depth and the datasets that can be retrieved.

*Social Media Data*

Prior to the advent of online social media, a large collection of data is exclusive to big research facilities such as weather forecast stations, astronomical stations, and scientific laboratories (Dijk van Jan 2006). Social media networks offer useful utility in understanding modern society and how it functions (Miller et al. 2015). It was estimated that 2.46 billion users will be connected in 2020, amounting to one-third of the global population.[7] Owing to the usefulness of the generated data, datafication[8], the continuous quest to turn every aspect of humans' lives into computerised data for competitive value (Cukier & Mayer-Schoenberger 2013), is being fueled by social media to supply commoditised data. Several domains have already recognised the crucial role of such data in improving productivity and gaining competitive advantages. Common use case examples include healthcare (Rojas et al. 2016, Yee et al. 2008), sport and entertainment (Davenport, 2014; Deloitte, 2014), politics (Contractor et al., 2015). The success of social media platforms has led to an increased interest in empirically testing various theories, making the platforms ideal for studying many aspects of social events. In terms of participants and data size, social media networks have profoundly transformed how various research works are being conducted, especially within the social sciences. Details about how researchers leverage theories, research constructs, and conceptual frameworks in relation to social media can be found in Ngai et al., (2015). Through *netnography*, researchers can systematically retrieve a huge amount of real-life observational data from different online social media platform's using traditional application programming interface (API) or a custom application.

---

[7] www.statista.com/topics/1164/social-networks
[8] see https://www.foreignaffairs.com/articles/2013-04-03/rise-big-data

*Twitter and Tweets* – Massive amounts of data can be easily obtained from platforms such as Twitter. Tweets, usually short text snippets, refer to the stream of posts users share on Twitter, and they enable longitudinal studies (Würschinger et al. 2016). A *tweet object* is a complex data structure, expressed in JavaScript Object-Notation (JSON) format, consisting of many extractable attributes that describe specific information about the *tweet* and the *account holder (the user)*. As a marked-up piece of text, the different fields in the tweet object define important characteristics of the tweet. The complexity of a tweet and its unstructured nature makes it difficult to process directly into a usable form, which requires a series of preprocessing before effective analysis can be conducted[9]. The stream of tweets differs from conventional stream texts in terms of posting rate, dynamism and flexibility; they are generated at a rapid rate and tend to be highly dynamic (Guille and Favre, 2015; Chakraborty et al., 2016).

**Relevant Work**

Since the outbreak, various stakeholders have been actively battling with the virus causing the pandemic, i.e.~SARS-Cov-2. To this end, researchers have been curating and documenting various online datasets about COVID-19, especially from social media. This endeavour is crucial towards enriching existing ground-truth data that could be used to debunk myths and misinformation around the pandemic. A collection of relevant datasets is central to tackling emerging challenges and a driving force in the various research efforts interested in combating harmful infodemic.

The following review is based on the modalities of misinformation spread, which include text, video, voice and images. It is out of the scope of the chapter to belabour or dwell on the research associated with COVID-19. There is a comprehensive catalogue of COVID-19 datasets, spanning various topics, in the work of Latif et al. (2020). The focus in this section of the chapter is on text and image modalities facilitated via social media networks, which have transformed the way sociological research is being conducted by enabling useful utility in understanding modern society and how it functions. Within a short span of the outbreak, there have been a plethora of COVID-19-related studies covering various aspects of the pandemic. Many datasets can be obtained from social media for various purposes related to the pandemic, such as in crisis management during the outbreak.

An early report about the first outbreak in China is summarised[10] in the work of Wu and McGoogan (2020). The Wikipedia projects have been maintaining comprehensive documentation about relevant articles on COVID-19. Using a large collection of diverse datasets from online social media, the *infodemics* observatory project keeps track of the digital response related to the outbreak (Valle et al., 2020). In conjunction with numerous online social media platforms, the World Health Organisation is preventing the spread of misleading information related to the pandemic[11]. Moreover, some social media platforms have put measures in place to prevent potentially inimical content from spreading. For instance, Twitter's new feature of flagging posts and the dedicated application programming interface can be used to retrieve tweets related to COVID-19[12]. A useful analysis of the impact of COVID-19 and how stakeholders can effectively act can be found in the blog post of Tomas (2020). Similarly, the focus in the work of Desvars-Larrive et.

---

[9] see https://github.com/ijdutse/covid19-datasets for some relevant information about COVID-19 data preprocessing.

[10] The following blogpost also offers useful insights about the pandemic: https://medium.com/@tomaspueyo/coronavirus-act-today-or-people-will-die-f4d3d9cd99ca

[11] see https://www.who.int/publications/i/item/9789240010314%20/

[12] see https://developer.twitter.com/en/docs/labs/covid19-stream/overview

al. (2020) is on how governments have implemented nonpharmaceutical intervention strategies on tackling the outbreak. The authors present a comprehensive structured dataset of government interventions and their respective timelines of implementations. Since various intervention measures have been applied, such measures will offer additional vista to understanding the progression of the pandemic at various stages.

*Social stream and evolving collection* – Some useful collections of social media datasets consisting of tweets can be found in the work of Chen and Ferrara (2020), and Alqurashi et al. (2020). A preprint version of the data contributed in this chapter is given in Inuwa-Dutse and Korkontzelos (2020). The work of Zarei et al. (2020) presents a collection of multilingual COVID-19 Instagram dataset that could be used to study the propagation of misinformation related to the pandemic. Qazi et al. (2020) present a large-scale multilingual tweet on the pandemic; the collection is composed of geolocation information that could be used for location-specific analysis of COVID-19 related issues. The work of Haouari (2020) presents COVID-19 dataset from Twitter based on the Arabic language. The work of Wang et al. (2020) presents an evolving large collection of diverse datasets from multiple sources/scientific research papers on COVID-19. Owing to its diversity, the collection is suitable for text mining and discovery systems related to the pandemic. Also, the work of Banda et al. (2020) contributes an evolving diverse dataset – biomedical, biological, and epidemiological – that captures social dynamics about the pandemic.

*Datasets on misinformation* – Noting the high spread of misinformation and fake health news over the Internet, there exists a plethora of studies and datasets on the topic. To help in tackling fake news and its detection in health news, Shahi and Nandini (2020) present multilingual cross-domain datasets of fact-checked news articles on COVID-19. In Dai et al. (2020), the focus is also on datasets (FakeHealth) to support fake news detection tasks. The work of Chen et al. (2020) contributes multilingual datasets suitable for tracking COVID-19-related misinformation and unverified rumours on Twitter.

The work of Memon and Carley (2020) is focused on characterising conspiracy theories and fake information to help in identifying and debunking unfounded claims. Similarly, the focus in the work of Cinelli et al. (2020) is on understanding the diffusion of information about the outbreak using datasets from various online platforms – Twitter, Instagram, YouTube, Reddit and Gab. The authors were able to track the spread of information from questionable online sources. Moreover, the following are useful websites that offer fact-checking services associated with the pandemic. *Snopes*[13] is an independent publication body that offers verification of misinformation on various topics, and *Poynter*[14] is part of the International Fact-Checking Network that provides a useful resource to help in neutralising COVID-19-related misinformation. The *Poynter* platform also initiated the hashtags *#CoronaVirusFacts* and *#DatosCoronaVirus* to gather misinformation about coronavirus.

## DATASETS AND RELEVANT RESOURCES

With the growing number of pandemic-related misleading information, another frontier of challenge in the fight against the virus is open. Figure 3 shows the focused areas in the fight against the COVID-19 pandemic. The fight against the virus revolves around preventive (such as public enlightenment about the standard practice to prevent spread) and curative. The best approach is to avoid endangering the public to be exposed to the virus. Of interest to most researchers, especially within the computer science research community, is the need to neutralise the negative impacts of infodemic associated with the pandemic. Infodemic could lead to consuming misleading information that could endanger the public. One approach to achieving such goals is through a useful data collection that could be used to debunk myths and misinformation around the pandemic. Thus, it is crucial to combat the pandemic from all angles using the right set of datasets.

---

[13] See https://www.snopes.com/
[14] See https://www.poynter.org/covid-19-poynter-resources/

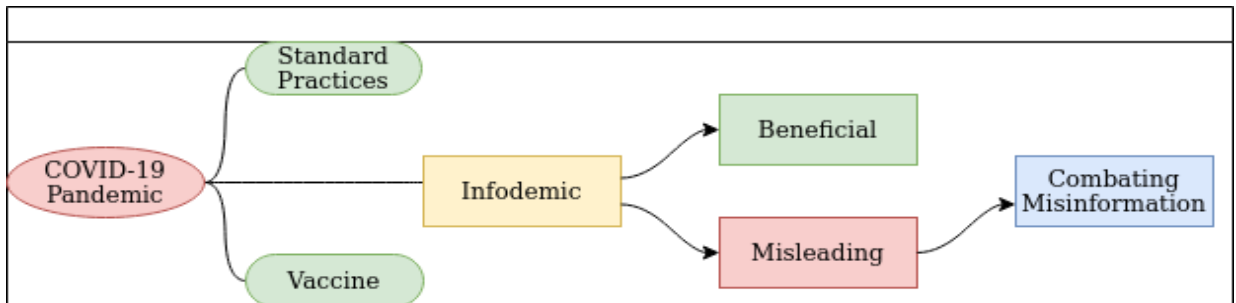

*Figure 3. The figure shows a high-level view of focused areas in the fight against the pandemic. Infodemic could lead to consuming misleading information that could endanger the public. Thus, it is crucial to combat the pandemic from all angles.*

To support the fight against the spread of misinformation and rumours, the following section describes the relevant datasets about COVID-19, how to retrieve the data contributed in the chapter, and a discussion on its potential utility. The collection consists of three categories of Twitter data, information about standard practices from credible sources and a chronicle of global situation reports from WHO[15]. Regarding data from Twitter, a description of how to retrieve the hydrated version of the data and some research problems that could be addressed using the data are given.

**Online Data Sources and Curation**

As discussed earlier in the Background section, the advent of social media has opened a new window of obtaining a huge number of diverse research datasets across different disciplines – engineering, medicine, sociology, computer science, etc. This section is concerned with a description of how to obtain and curate relevant datasets, notably from Twitter social network. Part of Table 1 at the end of the chapter provides an overview of useful tools for retrieving data from the respective social media platforms[16]. Figure 4 shows a basic collection pipeline of the datasets.

*Data Collection and Cleaning*

Platforms such as Twitter offer a useful avenue to retrieve a huge amount of data on a variety of topics using relevant keywords or search terms. The use of keywords plays a central role in identifying the most useful data and relevant stakeholders as the basis for data collection. The set of datasets presented in this chapter is in response to the growing scepticism, misinformation and myths surrounding the pandemic. Thus, terms that are associated with such myths have been used to collect the data, mostly from accounts that openly dismisses COVID-19 related information as put forward by credible sources such as the WHO. For a more effective result, it will be helpful to design the collection so that the data can be classified based on whether the collection is from *specific* or *dedicated* accounts or *random* or *miscellaneous* accounts via Twitter's API. The account-based collection could be from verified or unverified accounts on Twitter and the random set from a generic collection of daily tweets on diverse topics. These are needed to provide a wider context on the prevailing topic. Because a tweet associated with a hashtag offers a high-level filter and helps in data curation, the collection can be based on some specific hashtags.

---

[15] The datasets described in the section is available on arXiv preprint at https://arxiv.org/abs/2007.09703 (Inuwa-Dutse and Korkontzelos, 2020)

[16] See https://blogs.lse.ac.uk/impactofsocialsciences/2019/06/18/using-twitter-as-a-data-source-an-overview-of-social-media-research-tools-2019/ for a more comprehensive list of relevant tools.

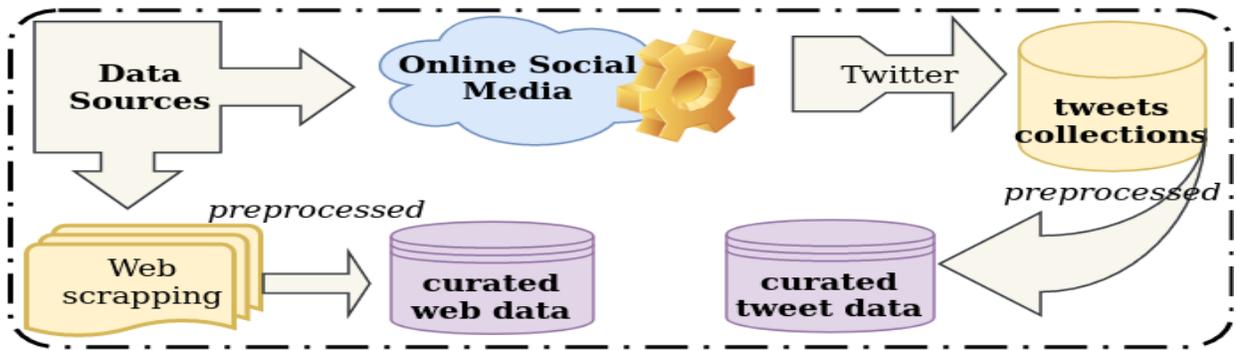

*Figure 4. A schematic illustration of the data collection pipeline.*

**Tweet-based collection -** All the tweet-based collections have been collected using Twitter's API and consists of three categories from accounts that have been monitored for five weeks (March 23 to May 13, 2020). Sometimes it is better to retrieve the whole tweet-object, a complex data object composed of numerous descriptive fields, instead of selected fields using tools such as Tweepy (see Table 1) because it enables the extraction of variables that could be used for further analysis.

**Non-tweet collection -** With the availability of many credible sources debunking misleading narratives about COVID-19, it will be beneficial to have a large collection of curated datasets. Consequently, the non-tweet group consists of information about standard practices from reliable sources and a chronicle of global situation reports on the pandemic. Bodies such as the World Health Organisation and nationally recognised institutions will be instrumental in providing rich and informative material for robust factual analysis. This will enable researchers to find responses to a wide range of questions related to the pandemic for a broader comparison and evaluation.

**Data cleaning –** To support effective longitudinal and exploratory analyses of the data, some crucial preprocessing steps are required. Basic forms of which include tokenisation, stopwords removal and text formatting (involving expanding contracted terms and lemmatisation). Once the data cleaning stage is completed, data analysis proceeds with a descriptive analysis to better understand the data before delving into the detailed study. Time-series analysis is crucial in revealing interesting patterns and useful insight.

## UTILITY OF PANDEMIC-RELATED DATA

The quest to turn every aspect of human life into computerised data for competitive value is rapidly growing. Depending on the interest and goal of the study, data from social media platforms can be used to conduct studies along the following dimensions: (1) Textual or multimedia data analysis: the prevalence of multimedia data (e.g., text, audio, video and graphics) enables various studies such as content and discourse analyses for many reasons (2) the second dimension is graph-based analysis, which relies on structural analysis to identify the underlying structure of relationships at various levels of granularity in social networks. Depending on which aspect or dimension is chosen, techniques or methodologies based on machine learning or deep learning can be applied to process or solve the problem at hand.

With data from online social media, there exist many useful theories, constructs, and conceptual frameworks to utilise for further investigation; the work of Ngai et al. (2015) offers more insight into the subject. Of interest to this study is to highlight areas where data from online platforms can be used to manage pandemic-related challenges in this age of hyperconnectivity. Potential problems to be addressed can be around pandemic outbreak detection and management, pandemic assessment, contingency planning, early detection or alert system for disease outbreak from online social media platforms, and modelling spread of outbreak at various levels.

### Detection of Spurious and Misleading Content

One of the reasons why online social media platforms are very popular with the public has to do with the ability for users to simultaneously generate and consume content leading to various forms of information – fads, opinions, breaking news. This reason also contributes to the increasing number of uncensored posts on various social phenomena (Inuwa-Dutse et al., 2018), partly due to their short size and the speed of communication. Among the repercussions of the increasing volume of information (relevant and irrelevant) on the pandemic is the tendency to create a sense of bewilderment on the part of the public concerning what preventive measures to take, and which piece of information to believe. There exist various misinformation and conspiracy sources capable of misleading the public regarding the COVID-19 pandemic. Demand for online-based services is at its peak during the lockdown, thus exposing the populace to various vulnerabilities. Despite the measures taken by social media platforms to curtail irrelevant content, many sources of misleading information and rumours still exist. As such, it is crucial to understand how online misleading content propagate and study how to optimise methods that favour the dominance of relevant content over irrelevant ones. A comprehensive repository of both validates and spurious datasets on pandemic will facilitate the authentication of the veracity of a given piece of information on the subject.

**Understanding Pockets of Outbreak**

Because users can share information about virtually anything, social media platforms are ideal for conducting useful studies. For instance, the data can help in informing what action to take that will prevent the occurrence or ramp up containment measures in a given locale. For an area not hit by the pandemic, mitigation measures and scenarios can be systematically categorised as *pre-outbreak*, *in-outbreak*, and *post-outbreak* to analyse situations and answer some beneficial questions. With the right data, some of the following crucial analyses, not requiring complex modelling, can be achieved (1) determine the number of cases – susceptible, infected and death (2) analyse the impact of the estimation (3) prioritize what course of action to take based on the prevailing situation and identify the most affected areas or groups. In terms of contact tracing, it will be interesting to ask the question about how possible it would be to trace susceptible cases using social media information. A basic strategy is to utilise self-reporting information about the relevant incidence, e.g., being in contact with an infected individual. As a result, the community will be proactive in handling any eventuality related to the pandemic because the level of preparedness will be improved significantly.

**Actionable Areas**

The infamous pandemic-induced lockdown has its many tolls across various sectors – at individual and societal levels. Thus, it is crucial to study its impact on mental health since people have been confined indoors, usually without jobs and momentous apprehension about eventualities. With self-reporting posts, triggers for viral infection and transmission within society can be studied to understand which of the imposed measures are more effective. Insights into these aspects will inform the best strategy to adopt in reaching out to the intended audience. Moreover, the study will help mitigate the 'scarring' effect of the pandemic and possible infrastructure damages as seen during the onset of the pandemic, which has been triggered by baseless claims, e.g., that 5G causes COVID-19.

**Flattening the Curves**

*The infection curve* – COVID-19, akin to illnesses such as diabetes, seems to require prolong precautionary measures to observe in managing it. Knowledge about the impact of COVID-19 is evolving, and its long-term effect is yet to be fully established. Thus, it is pertinent to incorporate diverse information sources from social media to help in offering a useful and holistic prevention pathway that is better equipped to tackle resurgences and other related eventualities. This endeavour is crucial noting how the healthcare facilities have been overwhelmed during the peak of the pandemic. A proactive approach will help towards improving the capacity of the health care service to fight the pandemic and effectively respond in allocating relevant resources intelligently and lessen future challenges. Accordingly, the following problems will be worthy of investigation.

- COVID-19 journey via online social media platforms: By leveraging self-reporting information on pandemic-related aspects such as lockdown, contact tracing, and public perceptions will offer relevant signals about recovery pathways. To complement the existing trusted public health information sources, harnessing information from online social media to address or examine the trajectory of the journey and its consequences at various stages – during the pandemic (pre-vaccine and post-vaccine) and after the pandemic (recovered individuals during the pandemic and post-vaccine recovery).

*The infodemic curve -* One of the challenges of managing pandemics during the age of online social media is the prevention of misinformation. With the prevailing situation, as the virus propagates, so is fake news or misinformation about it; hence it is equally important to simultaneously curtail propagation of the virus and its associated negative infodemic. A simple strategy that will go a long way in mitigating the harmful effect of negative infodemic will require each recipient of social media post to ascertain the veracity of seemingly problematic or controversial information before amplifying. Figure 5 shows a visual illustration from WHO where a simple verification process will prevent further spread. Abiding by this simple illustration will go a long way in curtailing the menace of misleading information, especially during the critical time of the pandemic Among other benefits, COVID-19 datasets could be used for various purposes such as the development of useful NLP-based tools for real-time digital disease surveillance from online news streams; to help identify the proliferation of spurious content, and to develop a real-time monitoring system of health-related content to inform preemptive measures.

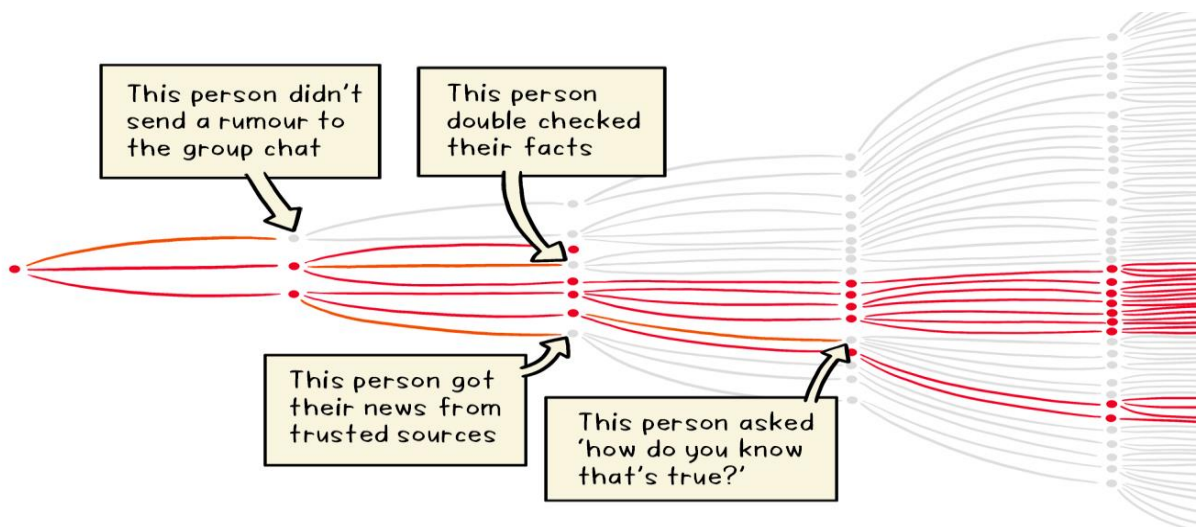

*Figure 5. A simple, albeit effective, means of filtering information is for each recipient to ascertain the veracity of the content/post, Source: ©WHO, 2020.*

**Community Detection and Sentiment Analysis**

**Community detection -** Networks are characterised by a certain degree of organisations in which groups of nodes form tightly connected units as communities. In the social science domain, sociometry is a means to measure or study social relationships between people (Wasserman & Faust 1994). Communities represent functional entities which reflect the topological relationships between elements of the underlying network (Newman 2006). Social network analysis is useful in revealing the dynamism of many forms of social relationships at various levels. Noting the level of resistance and acceptance regarding COVID-19, a high-level clustering of users could potentially unveil the distribution of users for reasons related to management, logistics, and containment of the outbreak.

**Sentiment Analysis –** Similarly, another useful problem to tackle is understanding users' perceptions about measures taken in managing the pandemic. For instance, it will be possible to evaluate lockdown policy to understand users' willingness to comply and the attitudinal change over time.

## CONCLUDING REMARK

The prevailing COVID-19 pandemic can be equated to illnesses that require prolong precautionary measures since the knowledge about its long-term impact is still evolving. Thus, it is pertinent to incorporate diverse information sources to help in tackling the pandemic crisis and offer a useful and holistic prevention pathway that is better equipped to tackle resurgences and other related eventualities. Since the outbreak, there has been a surge of studies investigating various aspects of the pandemic. Through online social media platforms, users are free to share personal information, which can be harnessed to mitigate some of the difficulties caused by the pandemic. Of interest to this chapter include studies centring on datasets from online social media platforms where the bulk of the public discourse happen. The datasets and relevant resources discussed in this chapter will play a significant role in tackling some of the pandemic-related challenges. Essentially, the following areas have been identified as crucial in the fight against the outbreak and could be studied using the datasets:

- curtailing the spread of inimical infodemic. This is needed because misleading information can have catastrophic consequences and hampers the fight about applying containment measures, which makes it pertinent to combat the pandemic from all possible fronts.
- among other benefits, the COVID-19 datasets, especially the online social stream, could be used for developing useful NLP-based tools for real-time digital disease surveillance, and to develop a real-time monitoring system of health-related content to inform preemptive measures.
- noting how the healthcare facilities have been overwhelmed during the peak of the COVID-19 pandemic, proactive strategy will help towards improving the capacity of the health care service to fight the pandemic and effectively respond in allocating relevant resources intelligently and lessening future challenges. The datasets and insights from the chapter will support studies interested in analysing the spread of fake and misleading content, evaluation of lockdown-related measures and tracking of public sentiment over time.
- the data will further enrich existing databases for debunking misinformation and fact-checking avenues, such as the International Fact-Checking Network.

Hopefully, the content of this chapter will support the computing community and policymakers in tackling present and future pandemics.

*Table 1. The table represents a sample of relevant tools to retrieve data from social media platforms. It is evident that Twitter dominates the list, which can be explained by the fact the data is more accessible than the rest of the platforms. Additionally, details about relevant resources for supporting studies on combating pandemic-related misinformation.*

| Tool | Description | URL |
|---|---|---|
| Brandwatch | Supported platforms include Twitter, Facebook, YouTube, Instagram, Sina Weibo, VK, QQ, Google+, Pinterest, Online blogs | https://www.brandwatch.com/ |
| Brand24 | Twitter, Facebook, Instagram, Blogs, Forums, Videp | https://brand24.com/features/#4 |
| Audiense | Twitter | https://audiense.com/ |
| IBM Bluemix | Twitter | https://www.ibm.com/cloud-computing/bluemix |
| Tweepy | Twitter | https://www.tweepy.org/ |
| Brand24 | Twitter, Facebook, Instagram, Blogs, Forums, Videp | https://brand24.com/features/#4 |
| Covid-19 stream | A dedicated API by Twitter to retrieve Covid-19-related tweets. | https://developer.twitter.com/en/docs/labs/covid19-stream/overview |

| Hydrator | A tool to return the full version of tweet's objects from Twitter. | https://github.com/docnow/hydrator |
|---|---|---|
| Worldometer | A website for keeping track of COVID-19 global cases. | https://www.worldometers.info/coronavirus/coronavirus-cases/ |
| | Framework for Managing Infodemics in Health Emergencies by WHO | https://www.who.int/publications/i/item/9789240010314%20/ |
| | A joint effort between WHO and the UK in documenting online misinformation | https://www.who.int/campaigns/connecting-the-world-to-combat-coronavirus/how-to-report-misinformation-online |
| Fact-checking | International Fact Checking Network. | https://www.poynter.org/ifcn/ |
| | Updates on COVID-19 diagnosis and treatment. | https://jamanetwork.com/journals/jama/pages/coronavirus-alert |